\documentclass[]{article}
\usepackage{authblk}
\usepackage{hyperref,aas_macros}
\usepackage[]{natbib}
\begin{document}

\title{Visions of Human Futures in Space and SETI}
\author{Jason T.\ Wright}
\affil{Department of Astronomy \& Astrophysics and \\ Center for Exoplanets and Habitable Worlds\\ 525 Davey 
Laboratory, The Pennsylvania State University, University Park, PA, 16802, USA, \href{mailto:astrowright@gmail.com}{astrowright@gmail.com}}
\affil{Visiting Associate Professor, Department of Astronomy  \\ Breakthrough Listen Laboratory \\  501 Campbell Hall \#3411, University of California, Berkeley, CA, 94720, USA }
\affil{PI, NASA Nexus for Exoplanet System Science}

\author{Michael P. Oman-Reagan}
\affil{Department of Anthropology, Memorial University, St. John's, NL A1C 5S7, Canada}

\maketitle 

\begin{abstract}

We discuss how visions for the futures of humanity in space and SETI are intertwined, and are shaped by prior work in the fields and by science fiction. This appears in the language used in the fields, and in the sometimes implicit assumptions made in discussions of them. We give examples from articulations of the so-called Fermi Paradox, discussions of the settlement of the Solar System (in the near future) and the Galaxy (in the far future), and METI. We argue that science fiction, especially the campy variety, is a significant contributor to the ``giggle factor'' that hinders serious discussion and funding for SETI and Solar System settlement projects. We argue that humanity's long-term future in space will be shaped by our short-term visions for who goes there and how. Because of the way they entered the fields, we recommend avoiding the term ``colony'' and its cognates when discussing the settlement of space, as well as other terms with similar pedigrees. We offer examples of science fiction and other writing that broaden and challenge our visions of human futures in space and SETI. In an appendix, we use an analogy with the well-funded and relatively uncontroversial searches for the dark matter particle to argue that SETI's lack of funding in the national science portfolio is primarily a problem of perception, not inherent merit. 


\end{abstract}


\section{Visions of SETI and Human Spaceflight}

The topics of SETI, human spaceflight, and humanity's long-term futures are intertwined. As we engage with outer space we bring history and culture with us and it becomes a ``cultural landscape'' \citep{gorman_cultural_2009}, a place that shapes and is in turn shaped by culture. Steven J. Dick\footnote{Dick is a historian of science, astronomer, former NASA Chief Historian, and former chair of astrobiology at the Library of Congress.} has called for a ``systematic approach'' applying anthropology to SETI and finds evidence in historical collaborations that this would be beneficial to both fields \citep{dick_anthropology_2006}. As a collaboration between an astronomer and an anthropologist, this paper draws on both interdisciplinary SETI work and anthropology to discuss factors shaping the cultural landscapes of space with special attention to ways we imagine and talk about possible futures through science fiction.

Human spaceflight, extraterrestrial intelligence, and the distant future are commonly completely blended in science fiction. Visions of humans traveling among the stars and encountering alien life are perhaps the quintessential science fiction trope. At first glance, it would seem that the endeavors of SETI and human spaceflight are generally more separated than this: the former has historically comprised relatively small, mostly privately-funded efforts concentrated among radio astronomers and a few others; the latter has comprised major governmental efforts by the United States, the former Soviet Union, Russia, and their partner nations for decades.\footnote{More recently China and India have begun human spaceflight programs, and private enterprises in the US such as SpaceX have announced their own ambitions.}

Here we consider the more ambitious projects for human spaceflight of the sort aspired to by many governments but never (so far) seriously attempted: the permanent settlement of the Solar System and beyond by humans. Visions of space settlement today often echo the narratives used during the dawn of the space age and the decades following, from early obsessions with space futures in the 1960s to the later focus on overcoming economic, social, and planetary limits in the 1970s \citep{mccray_visioneers:_2012}. At the same time the space race was beginning, the landmark paper ``Searching for Interstellar Communication'' was published in Nature \citep{cocconi_searching_1959} and Frank Drake performed his pioneering Project Ozma experiment \citep{OZMA}. Following popular interest and visions of space habitats in the 1970s, the SETI Institute was founded in  1984.\footnote{For more on the cultural history of SETI see \cite{capova_charming_2013}.} Today in the 2010s, alongside renewed talk of space settlement (lunar bases and settlements on Mars having taken the place of projects like the O'Neill cylinder \citep{ONeill} in the popular imagination) SETI efforts and interstellar spacecraft are now receiving some of their largest sustained funding to date through the Breakthrough Initiatives.\footnote{\url{http://breakthroughinitiatives.org}}

Human spaceflight and settlement efforts are connected to SETI by shared visions of the future in speculative fiction and imagination, and by the mathematics of the large spaces between the stars and the enormous age of the Galaxy.

For instance, astrophysicist Michael H. Hart calculated the time it would take a civilization to settle the Galaxy and found that even for conservative assumptions of space travel technology, a spacefaring community could have crossed the Galaxy many times by now.  This is the essence of the so-called \citep{FermiParadox} Fermi Paradox: if it's so ``easy'' to travel across interstellar space, then ``where is everybody?''  This line of reasoning has been used in attempts to discredit SETI as a serious endeavor, and many SETI theory papers dedicate themselves to rebutting it \citep[see][and citations therein for numerous examples]{GHAT1}.  

To illustrate this another way, a foundational concept in SETI, the Drake Equation, contains a final term, ``L'', representing the length of time a civilization may be transmitting signals, or the ``lifetime'' of a civilization \citep{DrakeEq,DrakeL}. A civilization that is destroyed (whether by self-destruction or some external cataclysm) will probably not be actively communicating, and so the opportunities for detecting alien civilizations are extremely sensitive to this term. But spaceflight enables a civilization to spread beyond its birthplace, inoculating it against many forms of annihilation and multiplying the number of potential transmission sites \citep[e.g.][]{GHAT1}. The success of SETI is thus strongly dependent on the prior success of interstellar spaceflight and settlement by alien species. Indeed, interstellar spacecraft {\it themselves} may be a signal detectable via SETI \citep{Yurtsever15,Loeb17}.

Contemplation of human spaceflight beyond Earth orbit is thus also tied to visions of humanity as an interplanetary or interstellar species---that is, our species' emergence as the sort of intelligence\footnote{In SETI jargon, the terms ``intelligence'' and ``intelligent species'' refer to life that is capable of engineering signals we are capable of detecting. It is well understood in the SETI community that this strictly functional definition of ``intelligence'' excludes many other forms of intelligence (including those that we have, have not, or perhaps cannot imagine).} that SETI seeks to detect elsewhere. The possible forms that alien intelligence and civilization might take are, in turn, influenced by science, speculation, fiction, and popular imagination. This often unacknowledged feedback loop of mutually producing visions of the future, built from both the stories we tell about alien civilizations and about our own history, results in conflicting visions for human futures in space which may included unintended bias. These conflicting visions are explicit in science fiction's treatment of extraterrestrial intelligence and a human interstellar diaspora; but in the actual practice of SETI and human spaceflight these conflicting visions are often implicit in the language or assumptions (often tacit) of practitioners. Our aim with the this paper is to provoke a more thorough recognition of this, and to inspire additional interdisciplinary research on this and related topics to the benefit of SETI and the future of human spaceflight.

\section{The Tropes and Language of SETI and Human Space Flight}

\subsection{Science Fiction's Influence}

Science informs much of science fiction, especially so-called ``hard'' science fiction that attempts to imagine possible technological futures with an emphasis on their possibility.\footnote{As opposed to ``space operas,'' ``soft'' science fiction, and other subgenres where alien life, space, and/or future technology are primarily plot devices or part of the setting.} Many foundational ideas in SETI and popular visions of humanity's futures in space come from scientists and engineers who also wrote hard science fiction, including Isaac Asimov\footnote{``Individual science fiction stories may seem as trivial as ever to the blinder critics and philosophers of today---but the core of science fiction, its essence, the concept around which it revolves, has become crucial to our salvation if we are to be saved at all.''\citep{asimov1981asimov}}, Arthur C.\ Clarke\footnote{``A critical---the adjective is important---reading of science fiction is essential training for anyone wishing to look more than ten years ahead.  The facts of the future can hardy be imagined {\it ab initio} by those who are unfamiliar with the fantasies of the past.'' \citep{clarke_profiles}} and Carl Sagan\footnote{Author of the influential 1994 book {\it Pale Blue Dot: A Vision of the Human Future in Space} \citep{PaleBlueDot} and the science fiction novel {\it Contact} \citep{contact} whose premise is the success of present-day SETI.}. 

Science fiction also informs science, inspiring scientists, shaping the history of science {\citep{Milburn:2010:MFS}, and playing a role in perceptions of science and the possible futures of humanity for both the public and practitioners \citep{lambourne_close_1990}.\footnote{Science fiction has influenced key figures in human spaceflight, from Wernher von Braun and Robert Goddard to Elon Musk, see \cite{mccray_visioneers:_2012}.} Since scientists, engineers, astronauts, and others involved in space projects were members of the lay public before their training, their ideas and attitudes are shaped by cultural products like science fiction, especially those who were or are fans of the genre. For example Mae Jemison, the first woman of color in space, was inspired to become an astronaut by actor Nichelle Nichols' portrayal of Lieutenant Uhura on the bridge of the {\it U.S.S.\ Enterprise} in the late 1960s television series {\it Star Trek} \citep{ST:TOS}.\footnote{Jemison paid the favor forward, appearing as an extra on {\it Star Trek: The Next Generation} \citep{ST:TNG:SecondChances}. Similarly, astronaut Sam Cristoforetti cited {\it Star Trek} as an inspiration---especially Captain Janeway from the {\it Voyager} series \citep{ST:Voyager}---and had herself photographed wearing a command uniform costume from the series while aboard the International Space Station.}
Presentations and discussions at space science conferences are often punctuated with references to science fiction intended for an audience of colleagues with an assumed familiarity. At an astrobiology science conference in April 2017, for example, one panel series was titled ``Seeking the Tricorder: Advanced Life Detection Tech'' in reference to the ubiquitous scientific instrument used in the {\it Star Trek} universe.\footnote{The Astrobiology Science Conference 2017 (AbSciCon), Mesa, Arizona, April 24–-28} Images of aliens from science fiction are regularly used in slides, scientists often bring up science fiction during interviews about their motivations for pursuing their work, and personal sections of professional web pages frequently list science fiction films, books, and games as an interest.

This has important consequences for SETI and the human exploration of space, because science fiction often uses ``aliens'' to build allegories of human relations rather than depicting more likely scenarios of first contact or realistic human spaceflight. This makes much of mainstream science fiction socially relevant and entertaining, but can have the effect of overshadowing more scientifically-grounded approaches to SETI and human spaceflight, as those fields inherit both the bias in the language and outdated assumptions of classic science fiction narratives.\footnote{Sometimes reporting about scientific research is transformed, via yellow journalism, into mediocre science fiction, as in the example of a paper by one of us \citep{PITS}. This paper's tacit premise was the {\it lack} of any evidence for ancient technological species in the Solar System, but sensationalist reporting claimed, for instance, that the author ``believes the aliens either lived on Earth, Venus or Mars billions of years ago.''  ``Have ALIENS lived on Earth before? Ancient `technological species' may have existed on our planet billions of years before humans, scientist claims'' {\it Daily Mail} Online April 25, 2017 by Shivali Best \url{ http://www.dailymail.co.uk/sciencetech/article-4443924/Ancient-technological-species-existed-Earth.html} Retrieved May 18, 2017.}

Consider, for instance, the clich\'e of a flying saucer landing on the White House lawn, the ``Mars Needs Women'' trope,\footnote{See {\it TV Tropes}: ``Mars Needs Women'' for examples of the trope and an excellent summary of its origins and relatives (\url{http://tvtropes.org/pmwiki/pmwiki.php/Main/MarsNeedsWomen}. Retrieved January 24, 2017)} and the tendency for first contact scenarios to almost universally involve white men as ambassadors of Earth. These all reflect the cultural biases of the creators and producers of these stories.\footnote{To say nothing of the outright racism and sexism, for instance in much of Robert Heinlein's work \citep[{\it Farnham's Freehold}, for example,][]{heinlein_freehold}.}  Even {\it Star Trek}---groundbreaking in its characters' ethnic and racial diversity---had white men in the top three roles (even the alien on the crew---Spock---was a white man!).  Beyond science fiction's role, SETI and human spaceflight build on bodies of work situated within a broader cultural history of inequality, and so inherit the perspectives and biases of their previous practitioners and society. This may be especially true of SETI, where much of the earlier work was done by a small group of people with a limited range of cultural and academic backgrounds, primarily Euro-American white men (with notable exceptions, e.g.\ Jill Tarter). 

\subsection{Colonial Inheritances of Imagined Futures}

Michael H. Hart's original calculation for the timescale of the settlement of the galaxy is usually discussed in terms of the ``colonization'' of the Galaxy. A central assumption in his argument is that if alien civilizations exist, ``they would eventually have achieved space travel and would have eventually explored and colonized the Galaxy, as we have explored and colonized the Earth.''\citep{hart75} This word choice evokes the colonial expansion of European powers around the world, and in particular uses ``we'' to refer to the explorers and the colonizers of Earth, but not the explored or the colonized. One might consider such a reading ungenerous, but given that Hart is a self-described ``white separatist'' \citep[][p.201]{Hart_White_Separatism} this connotation is probably not an accident. 

Prior to this, in an influential 1962 collection of essays, {\it Profiles of the Future}, Arthur C.\ Clarke discussed his vision for humanity's spread into space.  He did not imagine that Earth would rule over colonies: on the contrary he argued that the great distances of space would mean that ``the star-borne colonies of the future will be independent, whether they wish it or not.''  He imagined a future dystopia in which Earth had extreme population pressure\footnote{Regarding the consequences of not decreasing our birthrate, he wrote ``compulsory abortion and infanticide, and anti-heterosexual legislation---with its reverse---may be some of the milder experiments'' used to battle overpopulation.} which would drive some to colonize space in hopes of finding a better life.  On the question of who these colonists would be he wrote:

\begin{quote}
The age of mass colonization has gone forever. Space has room for many things, but not for ``your tired, your poor, your huddled masses yearning to breathe free.'' Any statue of liberty on Martian soil will have inscribed upon its base ``Give me your nuclear physicists, your chemical engineers, your biologists and mathematicians.''  The immigrants of the twenty-first century will have much more in common with those of the seventeenth century than the nineteenth. For the {\it Mayflower}, it is worth remembering, was loaded to the scuppers with eggheads.  \citep{clarke_profiles}
\end{quote}

Even granting Clarke's narrowest point---that astronauts will generally be few in number, highly trained, and hyper-competent---Clarke is still implying that the tired, poor masses greeted by Lady Liberty on their way to Ellis Island were mutually exclusive from scientists, mathematicians, and engineers. Clarke's myth also forgets the indigenous people who helped make the European settlements successful;\footnote{Over half of the {\it Mayflower}'s ``eggheads'' died in the first six months of their stay at Plymouth Colony; the colonists were taught how to farm and hunt indigenous foods by the Wampanoag.} the labor of enslaved people used to build the United States; the scientific and technological contributions they and their descendants made; and the immigrants (e.g.\ Nikola Tesla) and refugees (e.g.\ Albert Einstein) who brought both a yearning to breathe free and scientific competence.\footnote{31 of the last 78 Nobel Prizes in Physics, Chemistry, and Medicine awarded to Americans went to immigrants. National Foundation for American Policy, October 2016 Brief: ``Immigrants and Nobel Prizes'', \url{http://nfap.com/wp-content/uploads/2016/10/Immigrants-and-Nobel-Prizes.NFAP-Policy-Brief.October-2016.pdf}. Retrieved February 5, 2017}

In the examples above from Hart and Clarke, both authors write as members of a Euro-American civilization that has explored and colonized Earth and imagines it has bestowed comfort, culture, science, and technology on the rest of the planet's inhabitants over the past few centuries; not as members of a planetary human species that explored and settled nearly every habitable corner of the planet thousands of years ago. The former perspective is colonialist,\footnote{Clarke's writing, intentionally or not, is steeped in colonialism well beyond his invocation of the {\it Mayflower} above. {\it Childhood's End} \citep{clarke_childhood} can be read (spoiler alert) as a colonialist fable (if a somewhat ambivalent one), with humankind as the colonized people whose culture is annihilated as {\it their children} reject their heritage and accept membership in a superior culture.} and this language's early use shaped perceptions of SETI and human spaceflight as part of a ``linguistic ideology'' \citep[e.g.][]{kroskrity_language_2000} which, in turn, can shape and limit research and scholarship in those fields.

Well beyond Clarke's writing, colonialist perspectives have been quite common in science fiction from its earliest examples \citep{grewell2001colonizing}. Indeed, science fiction often imagines social and technological aspects of aliens to mirror human concepts of progress, with various species, civilizations, and worlds placed on a continuum from ``primitive'' and violent to ``advanced'' and enlightened.\footnote{This tendency is deliciously lampooned in {\it Mars Attacks!} \citep{burton_mars_1996} by the character Dr.\ Donald Kessler who assures the President of the United States regarding the comically violent and malevolent Martians: ``We know they're extremely advanced technologically, which suggests---very rightfully so---that they're peaceful. An advanced civilization, by definition, is not barbaric.''}  Such unilinear and hierarchical models of cultural change are holdovers from 19th century {\it progressivist social evolutionism} and the belief that European culture was the pinnacle of human civilization.  These models have since been abandoned by the disciplines that constructed them (early social sciences) and yet continue to appear in popular descriptions of both human futures and speculative alien civilizations.  This is mirrored in SETI in arguments that alien intelligence may be undetected so far because they all inevitably transcend beyond material form, or destroy themselves before their morality catches up with their technology.\footnote{In addition to claiming that all alien species will follow a similar social evolutionary path, these arguments also imply any given alien species can be characterized as a single ``civilization'' that is so homogeneous that the proffered explanation for our non-detection of any of them will apply to all members of their species, forever. They are thus examples of the ``monocultural fallacy'' \citep{GHAT1}.}

\subsection{Frontier Ideologies in Commercial Spaceflight}

Since the Space Shuttle program was retired in 2011, the U.S. space agency NASA has turned over, or is planning to turn over, most space transportation to private corporations and the ``commercial crew'' program. As venture capitalist space entrepreneurs and aerospace contractors compete to profit from space exploration, there are increasingly conflicting visions for human futures in outer space. Narratives of military tactical dominance alongside “New Space” ventures like asteroid mining projects emphasize the defense, privatization, and commercialization of space and other worlds over exploration and scientific research, thus framing space as a resource-rich ``frontier'' in what amounts to a new era of ``colonization'' \citep[e.g.][]{redfield_half-life_2002,valentine_exit_2012}.

Astrobiologist David Grinspoon recounts attending a conference\footnote{``Is Mars Ours?'' Slate, January, 7 2004, \url{http://www.slate.com/articles/health_and_science/science/2004/01/is_mars_ours.html}. Retrieved January 25, 2017} during which one speaker described terraforming Mars as ``the manifest destiny of the human race'' with all of the connotations of the phrase fully intended, turning what had been a discussion about the near future of human spaceflight into a debate on the ethics of American imperialism.  Grinspoon notes that a better analogy for Martian settlement than the violent story of the American West (which was, after all, inhabited at the time it was colonized) are the first migrations of humans out of Africa, and especially the first people to arrive and settle in the Pacific and the Americas.\footnote{Anthropologist Ben Finney, for example, drew on his research about Polynesian people crossing the Pacific Ocean to reconsider human migration into space in terms of historical migrations on Earth, and also to reconsider past human migrations in light of the emergent human migration into space. \citep{vakoch_tale_2014}} 

Biologist Danielle N. Lee observes\footnote{``When discussing Humanity's next move to space, the language we use matters.'' The Urban Scientist, {\it Scientific American} March 26, 2015, \url{https://blogs.scientificamerican.com/urban-scientist/when-discussing-humanity-8217-s-next-move-to-space-the-language-we-use-matters/}.  Retrieved January 30, 2017} how the language SpaceX CEO Elon Musk uses to talk about his ambitions for a privately funded Mars colony, echoing Clarke's vision, also perpetuates the perception of space flight as a way for a  (presumably rich) few to escape misery while the rest of us are ``stuck here on Earth.''\footnote{Quoting a summary of Musk's comments, {\it StarTalk Radio} Season 6, Episode 10,  \url{https://www.startalkradio.net/show/the-future-of-humanity-with-elon-musk/}. Retrieved February 2, 2017.}  She asks: whose ``version of humanity is being targeted for saving?,'' and connects this to work by Linda Billings, who documented the history of the language of justifications for the American space program.  \citet{billings_societal} identifies two competing narratives: the dominant one, ``a story of American exceptionalism that justifies unilateral action and the globalization of American capitalist democracy and material progress,'' and a competing narrative of ``utopian ideas of collective progress and a spiritual humbling of the self.''\footnote{Internal quotation marks and references omitted.}

\subsection{Tropes in the METI Debate}

The most glaring examples of conflicting visions of possible futures in SETI are in debates about Messaging to Extraterrestrial Intelligence (or METI, also known as ``active-SETI''). Instead of only passively searching for evidence of extraterrestrial intelligence, METI seeks to provoke a response, most commonly via powerful, deliberate signals that are far more likely to be detected by extraterrestrial life than our leaked radio emission.\footnote{Other forms of METI include the {\it Pioneer} plaques and the {\it Voyager} records---physical objects on interstellar spacecraft that contain messages.}

The most famous (and probably most powerful) attempt at METI was the 1974 ``Arecibo message'' composed by Frank Drake and others, and sent in the direction of M13, a cluster of about 300,000 stars around 21,000 light-years from Earth. Since then such efforts have been the subject of significant debate.  The two most recent efforts are those of METI International, founded by Douglas Vakoch,\footnote{Psychologist and former director of Interstellar Message Composition at the SETI Institute.} which seeks (among other goals) to develop and implement a METI program at the Arecibo radio telescope and other locations; and the Breakthrough Message initiative\footnote{\url{https://breakthroughinitiatives.org/Initiative/2}.  Retrieved January 25, 2017} which seeks to ``encourage debate about how and what to communicate with possible intelligent beings beyond earth.''   

The primary ethical concern is voiced most diplomatically in an open letter from members of the SETI community\footnote{``Regarding Messaging to Extraterrestrial Intelligence (METI) / Active Searches for Extraterrestrial Intelligence (Active SETI)'' \url{http://setiathome.berkeley.edu/meti_statement_0.html} Retrieved January 25, 2017. \label{Letter}} and most acerbically by \citet{Gertz_METI}, who concluded ``METI is unwise, unscientific, potentially catastrophic, and unethical'' because it seeks to attract the attention of potentially hostile or otherwise dangerous alien civilizations.  As such, some argue, METI would represent an existential threat to all life on earth.

Attitudes towards METI are varied, and generally reflect one's prior on the benefit or harm that would come from an alien intelligence noticing us.\footnote{One's ``prior'' being roughly a Bayesian {\it a priori} probability distribution; that is, one's internal, subjective sense of the relative likelihoods of various degrees of harm or benefit.  Also relevant is one's prior on the marginal likelihood that METI will result in us being discovered: one may feel, for instance, that METI efforts may make no difference, perhaps because our technosignatures are already obvious to technologically advanced civilizations, or because there are no alien civilizations, or because they will not care enough to look for us.}   The prospect of both great benefit and great harm exist \citep{Baum11}, and both have been explored in science fiction.  The prospect for beneficial contact is illustrated in work such as {\it Close Encounters of the Third Kind} \citep{spielberg_close_1977} and {\it Contact} \citep{contact}, while the prospect for harm is perhaps a more common trope,\footnote{A third trope appears in films like {\it District 9} \citep{blomkamp_district_2009} and {\it Alien Nation} \citep{baker_alien_1988} in which aliens arrive on Earth as refugees and are the victims of human prejudice and cruelty. These are rare, however, and from the {\it Alien} film series to popular video games, the extraterrestrial as threatening monster is ubiquitous.} for instance as illustrated in {\it Independence Day} \citep{emmerich_independence_1996} and {\it Berserker} \citep{Berserker} \citep[and going back at least as far as {\it The War of the Worlds,}][]{wells1898war}.\footnote{Earlier examples of first contact in writing include Voltaire's {\it Microm\'egas} (1752) and Lucian's {\it True History} (written between 125-180 CE).}

Much of this discussion in the academic literature is based upon analogies with human culture just as it often is in science fiction narratives. While it is all but certain that intelligent alien civilizations will be much older than humanity,\footnote{In brief, {\it H. sapiens} is a few orders of magnitude younger than the typical ages of stars and planets in the Milky Way, and so unless intelligent life arises practically simultaneously everywhere, we are either the first such life, or others are at least millions of years older than us \citep[cf.][]{Annis99a,cirkovic08}.} it is by analogy with popular ideas about human technological progress that we presume they will both be much more technologically advanced than us and what form that ``advancement'' will take.\footnote{``We know absolutely nothing about ET's intentions; however, its ability to do us harm, if it so wished, might be absolute.  We are only two thousand years more advanced than Rome, yet the best Roman legion could be annihilated by a modern army in mere minutes. Never mind two thousand years ago, Napoleon's armies of barely two hundred years ago would face the same fate.'' \citep{Gertz17}} Similarly for the perceived results of any interaction: the same ideologies underlying the desire and justification for ``colonizing'' space might also make some wary of contacting other intelligent life, lest humanity become the victims of the alien colonization they imagine would likely follow.\footnote{``If aliens visit us, the outcome would be much as when Columbus landed in America, which didn't turn out well for the Native Americans...advanced aliens would perhaps become nomads, looking to conquer and colonise whatever planets they can reach.'' Stephen Hawking, from a Discovery Channel documentary, quoted in ``Stephen Hawking takes a hard line on aliens'', {\it The Guardian} April 26, 2010 by Leo Hickman  \url{https://www.theguardian.com/commentisfree/2010/apr/26/stephen-hawking-issues-warning-on-aliens}. Retrieved February 11, 2017.} 

The opposite perspective may come from presuming that certain forms of social and ethical change would be an inevitable consequence of a civilization's survival over a long period of time,\footnote{``The idea of a civilization which has managed to survive far longer than we have...and the fact that that technology remains an aggressive one, to me, doesn't make sense...The pressure of long-term survival---of limiting population...I think requires that the evolutionary trends that ratcheted up our intelligence...continues to evolve into something that's cooperative and take on global scale problems.'' Jill Tarter, interview with Jessica Orwig in {\it Business Insider}, January 20, 2016, ``A world leading scientist on the search for extraterrestrials pointed out a flaw in Stephen Hawking's fear of finding intelligent aliens'' \url{http://www.businessinsider.com/jill-tarter-says-stephen-hawking-is-wrong-about-aliens-2016-1} Retrieved February 11, 2017.} or from a strong desire to {\it avoid} projecting human intentions onto alien life, and the presumption that violence and colonialist tendencies are uniquely human traits.  Both assumptions introduce problems for SETI research because they are built on analogies with popularized accounts of human history; visions of history which are already politicized and partial \citep{denning_impossible_2013}.

These two perspectives may also lead to divergent opinions on the necessity and practicality of human spaceflight and SETI. {\it Star Trek} and {\it Cosmos} \citep{sagan1985cosmos} envisioned humanity taking its place among the stars, perhaps as part of a Galactic community of civilizations (``One Voice in a Cosmic Fugue,'' to use Sagan's chapter title). From a colonialist perspective, however, ambitions for human spaceflight and SETI are antithetical: one seeks to realize humanity's dominion of the Galaxy, the other seeks to demonstrate that another form of life may have beaten us to it.\footnote{``Even if we never reach the stars by our own efforts, in the millions of years that lie ahead it is almost certain that the stars will come to us\ldots And when the first contact with the outer universe is made, one would like to think that Mankind played an active and not merely a passive role---that we were the discoverers, not the discovered.'' \citep{clarkeexploration}} This perspective might lead to championing the colonization of the Solar System and beyond, and motivate arguments against the existence of life elsewhere in the Galaxy, since such life might challenge often racialized and culturally specific ideas about human uniqueness and superiority.\footnote{For example, the book {\it Extraterrestrials: Where Are They?} \citep{zuckerman1995extraterrestrials}, which takes a skeptical look at the prospects for alien intelligence existing in the Galaxy. Its editors are the aforementioned ``white separatist'' Michael H. Hart, and the president of ``Californians for Population Stabilization,'' an anti-immigration group labeled a ``hate group'' by the Southern Poverty Law Center. ``California Today: The State's Hate Landscape'' {\it The New York Times}
February 22, 2017, by Mike McPhate. \url{https://www.nytimes.com/2017/02/22/us/california-today-hate-groups.html} Retrieved April 15, 2017.} 

\subsection{The ``Giggle Factor''}

	SETI, human spaceflight beyond the Moon, and related topics also suffer from a ``giggle factor.''  Scientists attempting to study issues like intelligent life in the universe, planetary protection for asteroid impacts\footnote{```Giggle factor' is no laughing matter to scientists'', USA Today,  March 11, 2003, by Eric J.\ Lyman, \url{http://www.ericjlyman.com/usagigglefactor.html}. Retrieved January 25, 2017}, and the development of structures and human settlements in space\footnote{Arthur C.\ Clarke said that ``the Space Elevator will be built about 50 years after everyone stops laughing,'' \citep{ClarkeLaughing} and attributed the quip to Arthur Kantrowitz regarding Kantrowitz's space laser propulsion system.} are often not taken seriously by the public, politicians, and their colleagues.  This has serious consequences for these fields.\footnote{And studies {\it of those fields}.  Sociologist Albert A.\ Harrison warns that ``sociologists whose activities can be linked to `little green men' risk ridicule and professional censure'' \citep{harrison_overcoming_2005}. An important distinction is the disciplinary acceptance of those who study SETI researchers but not those who engage in the research questions of SETI.
    From the 1970s to the 1990s NASA also supported social science work in SETI. For reviews of social science involvement in SETI see \citep{harrison_role_2000} and \citep{dick_anthropology_2006}.}
Most infamously, perhaps, was the ``Golden Fleece Award'' bestowed by U.S. Senator William Proxmire upon the peer-reviewed SETI research of NASA, chosen as an exemplar of frivolous spending of taxpayer dollars.  NASA attempted to maintain funding for SETI research by keeping a lower profile for the project, but the temptation for members of Congress to exploit the ``giggle factor'' for grandstanding purposes proved too great. In 1993 the last major NASA SETI effort was canceled by Congress, with Senator Richard Bryan boasting in a press release ``This hopefully will be the end of Martian hunting season at the taxpayer's expense''---although the program in question involved searches for interstellar radio signals, not an exploration of Mars  \citep{Garber99}.

All of the space sciences, from the study of exoplanets, to black holes or gravitational waves, respond to the challenges of producing knowledge about distant, sometimes speculative, objects and phenomena \citep{messeri_resonant_2017}. Today, astrobiology (the search for and study of the nature and origins of life in the Universe) is one of NASA's top research priorities, while SETI is all but absent from its research portfolio.\footnote{``While\dots[radio SETI] is not part of astrobiology, and is currently well-funded by private sources, it is reasonable for astrobiology to maintain strong ties to the SETI community.\dots Rather than argue for or against the likelihood of finding [signatures of technology], or attempt to describe specifically what such a signature would look like, we should be sure to include
it as a possible kind of interpretation we should consider as we begin to get data on the exoplanets.'' \citep[][Astrobiology Roadmap, p.\ 150]{NASA_Astrobiology_2015}} There is no scientific reason to search only for {\it unintelligent} alien life when {\it intelligent} alien life may be much easier to find, and so one significant reason for the disparity is the ``giggle factor'' and the risk of Congressional reaction.\footnote{Another is that the ``giggle'' is a nervous one---microbial ``slime'' on another planet does not threaten humanity's privileged place in the universe (philosophically or physically) and so the prospect of finding it may arouse less opposition.} 

It is true that many, such as Hart, have argued, based only on speculation and the fact that intelligent alien life is not obvious to us, that it {\it must not exist} \citep{tipler80,zuckerman1995extraterrestrials,Gott93}. While these arguments are often used as an {\it excuse} to exclude SETI from funding calls, such leaps of logic do not similarly exclude other speculative fields such as astrobiology and dark matter detection (see the Appendix for an exegesis of this point).\label{darkmatter} 

As \citet{Chyba05} noted:
\begin{quote}
Astro-physicists...spent decades studying and searching for black holes before accumulating today's compelling evidence that they exist. The same can be said for the search for room-temperature superconductors, proton decay, violations of special relativity, or for that matter the Higgs boson. Indeed, much of the most important and exciting research in astronomy and physics is concerned exactly with the study of objects or phenomena whose existence has not been demonstrated---and that may, in fact, turn out not to exist. In this sense astrobiology merely confronts what is a familiar, even commonplace situation in many of its sister sciences.\footnote{Internal references omitted.}
\end{quote}

Human spaceflight of the sort we are concerned with has similarly suffered from a lack of government support and ``giggle factor''.\footnote{In addition to the reasons discussed here and in the appendix, SETI and space settlement funding is also shaped by structural political economic forces, relations of power, and structures of authority, see \cite{redfield_half-life_2002}, \cite{valentine_exit_2012}, \cite{graeber_utopia_2015}, and  \cite{genovese_new_2017} for examples.} Aside from a brief, quixotic burst of effort during the Apollo era, most governmental ambitions for human spaceflight have been confined to low earth orbit in the form of space stations, the Soviet-era (now Russian) Soyuz capsules still in use, and the now-retired US Space Shuttle. Indeed, discussion of the only (apparently) seriously funded effort for a permanent human presence beyond Earth, the SpaceX aspirations for a Mars settlement, is pervaded by fancy, including talk of planetary apocalypse, tyrannical runaway artificial intelligences, and supervillians.\footnote{For instance, ``Elon Musk: Our Savior, The Supervillain: How one technology mogul's terror of extinction serves the human race,'' {\it Popular Science}, November 24, 2014, by Eric Sofge. \url{http://www.popsci.com/our-savior-supervillain}. Retrieved April 14, 2017. Musk has skillfully embraced this tendency to enhance his brand: ``Even Elon Musk knows he's a good supervillain candidate'' {\it The Washington Post} April 17, 2015, by Andrea Peterson  \url{https://www.washingtonpost.com/news/the-switch/wp/2015/04/17/even-elon-musk-knows-hes-a-good-supervillain-candidate/} Retrieved April 15, 2017.}

\subsection{Science Fiction's Role in the ``Giggle Factor''}

Many have argued that SETI's ``giggle factor'' is due in large part to ufology and reports of alien abductions \citep[e.g.][]{achenbach2003captured}. These fields were debunked decades ago and today lack scientific credibility except as subjects for studies of human behavior and culture \citep{dean_aliens_1998,battaglia_e.t._2006}, but remain closely associated with SETI in the minds of the public because of their similar subject matter. While there is certainly much true in this explanation, this sort of popular confusion is not unique to SETI: many topics in physics and astronomy have analogous associations with paranormal and New Age topics such as constellations, crystals, energy fields, and quantum phenomena.  So while SETI may always struggle with this association, it is hardly a sufficient explanation for the ``giggle factor.'' 

One factor for this particular struggle of SETI, and also human spaceflight, is likely the popular depictions of science, spaceflight, and alien life in science fiction.  While the genre has plenty of examples of high quality, certainly most of the genre by volume is rather poorly made. Even some classic examples of the form, such as {\it Star Trek}, suffer from many of the traits that have come to exemplify the form in all of its media: cheap production values, poorly written dialogue, ham-handed acting, unimaginative plots, and tissue-thin allegories and metaphors.\footnote{Recent {\it Star Trek} films \citep{ST:Reboot} focus on fighting, action, and explosions in contrast with {\it Star Trek} creator Gene Roddenberry's vision for the series: ``The whole show was an attempt to say that humanity will reach maturity and wisdom on the day that it begins not just to tolerate, but to take a special delight in differences in ideas and differences in life forms. If we cannot learn to actually enjoy those small differences, take a positive delight in those small differences between our own kind, here on this planet, then we do not deserve to go out into space and meet the diversity that is almost certainly out there.'' New York City, May 1976, Transcript from \url{http://www.niatu.net/transfictiontrek/download/gene-roddenberry-st-philosophy.pdf}. Retrieved February 5, 2017}  Many or most lists of the ``worst films of all time'' are dominated with science fiction films.\footnote{And their close cinematic cousins, fantasy and monster films. See, for instance, Wikipedia's list at \url{https://en.wikipedia.org/wiki/List_of_films_considered_the_worst}.  Retrieved January 25, 2017}

Thus, when the topic of alien life or human spaceflight to other planets comes up, the immediate association in the minds of the public (and many scientists) may be these popular depictions of the topics, which are predominantly campy,\footnote{Roughly in the \citet{SontagCamp} sense of ``naive'' ``failed seriousness.''} at best. From the first mention of the topic, then, the burden is on those trying to discuss SETI, large structures in space, or the human exploration of space to defend the seriousness of the endeavor, and to overcome this initial reaction which is often a snicker or ``giggle.'' Scorn for the ``geek'' culture often associated with science fiction fandom may also play a role, even among professional scientists and engineers, a question worth further investigation.

\section{Toward New Visions of Humanity's Futures}

\subsection{Changing the Language}

As we have seen, the terms, assumptions, and visions SETI and human spaceflight inherit come with historically situated exclusionary cultural baggage that we then carry into our visions of the future. Many of the implications of that baggage are not accidentally brought along, but intentionally preserved, and this language has real effects on scientists' and the public's perception of SETI and spaceflight. Regardless of their personal politics, it behooves practitioners to know the origins of the terms they use and visions for the future they hold, and examine the biases they bring, both to avoid unacknowledged bias in a field that requires an open mind, and to ensure that the field itself is not excluding voices and perspectives that will also help it thrive.

SETI, more than many disciplines, requires keeping an open mind, since it deals with possible forms, motivations, and natures of alien life. Anthropocentrism pervades the field as people, in the almost complete absence of data, project human values and tendencies onto hypothetical alien civilizations \citep[e.g.][]{denning_ten_2011}. A variety of cultural, experiential, and disciplinary backgrounds among SETI practitioners would help expose and correct many of the biases and assumptions, and bring fresh ideas for search methods to the field.\footnote{We can go even further: including our present-day interactions and relationships with other species in our analysis can help us remain open to possibilities of ``multispecies'' futures in space, and provide data for more informed speculation about what interactions with extraterrestrial life might be like \citep{herzing_profiling_2014, oman-reagan_social_2015}.}

So, too with human spaceflight. NASA has already deprecated the term ``(un)manned'' to refer to whether missions and vehicles were crewed by people (preferring ``human,'' ``(un)piloted,'' ``(un)crewed,'' ``robotic,'' etc.).\footnote{``Style Guide for NASA History Authors and Editors'' NASA History Program Office \url{https://history.nasa.gov/styleguide.html}. Retrieved February 5, 2017}  This requires care: Emily Lakdawalla\footnote{``Finding new language for space missions that fly without humans,'' {\it Snapshots from Space} blog, October 5, 2015, \url{http://www.planetary.org/blogs/emily-lakdawalla/2015/10050900-finding-new-language.html}. Retrieved February 3, 2017} notes that style guides for major media outlets, including {\it The New York Times} and the {\it Associated Press} nonetheless insist on ``(un)manned;'' that ``crewed'' is unfortunately a homophone for ``crude;'' and that humans can pilot robotic ships, while computers can pilot crewed ones. Nonetheless, this sort of shift in language matters: space archaeologist Alice Gorman\footnote{``How to avoid sexist language in space---Dr.\ Space Junk wields the red pen.'' {\it Space Age Archaeology} blog September 6, 2014,  \url{http://zoharesque.blogspot.ca/2014/09/how-to-avoid-sexist-language-in-space.html}. Retrieved February 3, 2017} notes
\begin{quote}
When you're a bloke, terms such as ``mankind'' automatically include you. You don’t have to think about it at all; you're already in there. Now we all know that these terms are {\it supposed} to also include women; but the reality is a bit different...women have to ``think themselves into'' such expressions, even if it happens at a subconscious level. 
\end{quote}

Of course, it may be impossible to choose language entirely free of unintended, biased, or otherwise problematic implications. Using ``settlement'' for ``colony'' might evoke Israeli settlements in the Middle East; and terms like ``migration'' and ``relocation'' have been used as euphemisms for dispossession and genocide.  Nonetheless, the term ``colonize'' is so loaded, and some of its earliest and most influential use in the field was so deliberate, that ``settle,'' ``migrate,'' and their cognates should be preferred today, much as they are used to describe the first human migrations across the globe.  Other terms to be reconsidered abound:  ``mankind's'' ``Manifest Destiny'' in space,  ``conquering'' space or other planets, and alien ``races'' all have similar pedigrees to ``colonize.''

So too with portrayals of astronauts and other future inhabitants of space. Our future will only represent a small subset of humanity unless perceptions of who goes into space are made broader than Clarke's, to include humanity in all our diversity.\footnote{Douglas Vakoch has applied the idea of including diversity to METI message composition \citep{vakoch_dialogic_1998}.} The universality in Carl Sagan's vision of Earth as the cradle of humanity is because his Pale Blue Dot is home not just to ``everyone you ever heard of'' \citep{PaleBlueDot}, but also everyone you've {\it never heard of}---the marginalized, oppressed, erased, and forgotten.

The diversity of the existing astronaut corps matters because governmental agencies and private space interests are preparing new craft and taking humans into space now. In late 2015, the entire crew of the International Space Station were male and members of their home country's ethnic majority, and no openly LGBTQ person has orbited the Earth.\footnote{The sexuality of Sally Ride, the first American woman in space, was private until her death.} On the other hand, for the first time in history, the NASA class of eight astronauts recruited in 2013 was 50 percent women, which has special relevance because they may be among the first humans to go to Mars.

What's more, the demographics of today's astronauts may have implications far into the future. Clarke's point about the independence of interstellar settlements will also apply to their cultural development \citep{GHAT1}.  Without low-latency interactions with Earth and each other, the various human settlements will undergo extreme cultural drift, similar to the way cultures grew apart on Earth in the era before global communication networks, but potentially more pronounced.\footnote{For examples in science fiction writing see the Hainish series \citep{le_guin_dispossessed_1974} and the Twenty Planets series \citep{gilman_dark_2015}, which both imagine human diaspora on many worlds becoming alien to one another thousands of generations after Earth seeded planets throughout interstellar space.} They will inevitably consider themselves distinct from the cultures of their homelands, but these new, distinct cultures will nonetheless inherit many aspects of their founders' cultures. This means that the choice of who these first settlers are---and the perspectives and values they bring with them---will have profound effects on the futures of humanity.\footnote{A point explored in the {\it Mars} trilogy \citep{RedMars}.}  As the dream of a human settlement on Mars drifts closer to becoming reality, the composition of its first inhabitants should become a topic of serious, thoughtful, and widespread discussion.

Competing visions of human futures in space are also apparent in current discussions about Earth's environment and climate change. As scientific exploration and human spaceflight transform the solar system from a cold and dark place ``out there'' into a place where humans live and work (such as on the International Space Station and in plans for Mars habitats), space becomes an extended part of Earth's environment \citep{olson_american_2010}. An ongoing effect of this is that ideas about who goes to space and what kind of environments they live and work in are imported back to Earth. Concepts like ``carrying capacity,'' for example, had their origins in astronautics but are now used to think about planetary habitability on Earth \citep{anker_ecological_2005}. These examples demonstrate that the ways we imagine and build human futures in space will continue to shape social, cultural, scientific, and environmental futures down here on Earth.

\subsection{Imagining Futures Otherwise}

If science fiction has played an important role in weighing down SETI, human spaceflight, and visions of humanity's futures with baggage, it can also serve as a corrective and an opening for better possible futures.

For instance, it is possible that the ascension of ``geek'' subcultures \citep[e.g.][]{lockhart_nerd/geek_2015,morgan_rise_2014} and the resulting proliferation of high-production-value and high-concept science fiction on television and in film will alleviate the ``giggle factor'' to some degree, especially among younger scientists. The recent film {\it Arrival} (2016) departed from the campy unscientific depictions of contact with extraterrestrial intelligence by incorporating scientific contributions from computing \footnote{``Analyzing and {Translating} an {Alien} {Language}: {\it Arrival}, {Logograms} and the {Wolfram} {Language}'', Wolfram Blog, {\url http://blog.wolfram.com/2017/01/31/analyzing-and-translating-an-alien-language-arrival-logograms-and-the-wolfram-language/}. Retrieved May 3, 2016.} and linguistics.\footnote{``For linguists, {\it Arrival} can't come soon enough'', {\it Science} \textbar\ AAAS, November 11, 2016, by Brice Russ. \url{http://www.sciencemag.org/news/2016/11/linguists-new-sci-fi-film-arrival-cant-come-soon-enough}. Accessed May 11, 2017.}  The high profile NASA is bringing to astrobiology along with serious discussion of human habitats on Mars by both NASA and private enterprises might also help to alleviate the ``giggle factor''.

Historian of science Colin Milburn makes a compelling argument that science fiction can function as a ``repository of modifiable futures'' in science \citep{Milburn:2010:MFS}. Some science fiction thus provides opportunities for fruitful speculation about extraterrestrial intelligence beyond classic anthropocentric assumptions and helps us to think about alternative possible futures \citep{collins_all_2008,oman-reagan_unfolding_2015}. The ``exercise of imagination,'' Ursula K.\ Le Guin reminds us, ``has the power to show that the way things are is not permanent, not universal, not necessary'' \citep{le_guin_wave_2004}.   Anthropologically informed science fiction addressing the social and cultural issues of space exploration and contact with extraterrestrial intelligence can also help scientists see the value of social scientific approaches to their research projects. Just as astrobiology draws on multiple disciplines to speculate about possible life elsewhere, speculative forms of anthropology can engage productively with science and fiction to imagine possible civilizations elsewhere and otherwise, both human and non-human.\footnote {Experimental forms of ethnographic writing that blend fiction and analysis to move beyond species and nature/culture boundaries offer one example of the forms this can take. For example \cite{tsing_strathern_2014} writes an ethnographic account from the perspective of a fungal spore to explore how we think about human and non-human social relations.} Considered in this light, practices of thinking alongside science fiction may allow scientists and authors to collectively imagine new approaches to SETI and better futures for humanity in space.\footnote{\citet{Haqq16} suggests a more radical, long-range approach: encouraging a human settlement on Mars to  develop culturally independently of Earth, in order to generate new moral and ecological perspectives on ``planetary citizenship''.}  The following are a few such examples (beware, spoilers ahead).

In Carolyn Ives Gilman's novel {\it Dark Orbit} \citep{gilman_dark_2015}, an interstellar ``exoethnologist'' \footnote{From exo-, external, and ethnology, the study of cultural differences and relationships between communities.} studies human-descended communities living entirely in darkness on an exoplanet who have evolved to rely only on sound instead of light-based sight, {\it seeing} physical space by listening. Gilman, who is also a historian at the Smithsonian, skillfully imagines profound changes in culture based on variation in planetary conditions, providing an example of informed speculation about the ways human migration into space may change our societies.

In Peter Watts' book about first contact, {\it Blindsight} \citep{watts_blindsight_2006}, extraterrestrials regard human use of language to communicate information about the individual self as an existential threat. In Stanis\l{}aw Lem's first contact novel {\it Solaris} \citep{stanislaw_lem_solaris_1970}, scientists spend decades studying an intelligent ocean on a distant world, but its vast alien consciousness is so different from theirs that they have no hope of understanding it. Imagining these sorts of radically different life and intelligence may help us do what \citet{cabrol_alien_2016} has suggested is necessary for future SETI research; to ``step out of our brains'' and move beyond the problem of searching only for ``other versions of ourselves.''

Science fiction written by women, Black people, Native people, and others often explicitly challenges the colonialism, sexism, and anthropocentrism\footnote{See, for example, Octavia E. Butler's ``Xenogenesis/Lilith's Brood'' books, Ursula K. Le Guin's ``The Word for World is Forest.''} that pervade both the genre as well as the human spaceflight industry.\footnote{For example see {\it Walking the Clouds:An Anthology of Indigenous Science Fiction} \citep{dillon_walking_2012} and {\it Octavia's Brood: Science Fiction Stories from Social Justice Movements} \citep{imarisha_octavias_2015}}  For instance, Black science fiction and Afrofuturism can challenge portrayals of American Black cultures in space and science-fiction as an absurdity.\footnote{Such a portrayal served as the central joke in the television sit-com {\it Homeboys in Outer Space} \citep{ehrich_van_lowe_homeboys_1996} For analysis of how the science fiction genre has represented Blackness see \cite{nama_black_2008} and \cite{thomas_dark_2000}.} Work by Octavia E. Butler, Samuel R. Delany, and others address the ``intersectionality'' \citep{crenshaw_demarginalizing_1989, crenshaw_mapping_1991} of race, gender, class, power, sexuality, and human futures in ``hard'' science fiction. The ``indigenous futurisms'' appearing in films by indigenous people offer alternative visions of first contact along with visions of anti-colonial futures in space which are grounded in history \citep{lempert_decolonizing_2014}, rather than trying to simply forget colonial pasts and inheritances.  Each of these examples offer novel ways of thinking about the problems of human adaptation to space and detecting extraterrestrial intelligence precisely because they challenge assumptions often made about what cultures and civilizations elsewhere might be like---namely, like a particular subset of humans today on Earth.

By turning to diverse, boundary-challenging works like these, SETI and human space exploration can gather inspiration for new ways of thinking about central questions of the disciplines and the benefits to science and society of diversity and inclusiveness.\footnote{The role of science fiction in scientifically informed speculation can also be understood in terms of ``speculative fabulation,'' a notion developed by Marlene Barr \citep{barr_feminist_1992} to examine women's writing as ``feminist fabulation'' and later expanded by Donna Haraway \citep{haraway_sf:_2013}. Speculative fabulation is a theoretical apparatus for analyzing modes of writing and imagining which can be applied to both science and fiction.}  Until more science fiction works like these pervade the public sphere and the disciplines better reflect human diversity, scientists and engineers should be careful to avoid reproducing the biases they inherit and do their best to ensure {\it all} of humanity and the diversity of life on Earth are included in their visions of possible human futures.

\appendix
\section*{Appendix: Analogy Between Direct Dark Matter Detection and SETI}
\label{appendix}

As noted in the \S\ref{darkmatter}, SETI suffers a large funding disparity with respect to similarly speculative fields, such as dark matter particle detection experiments. Indeed, the analogy between dark matter particle detection and SETI is quite good. In this appendix, we develop the analogy, as it sharpens our argument that the funding disparity is based on popular perceptions of the topic, not the inherent plausibility of the endeavor. 

We know from several lines of reasoning and observation that our standard models of physics and cosmology are missing an important piece. For instance, the motions of the gas within galaxies and the relative motions of galaxies themselves are not well described by the standard model of gravity acting on ordinary matter. Many arguments are made on ``naturalness'' and other mostly philosophical grounds\footnote{i.e.\ proposed solutions are generally favored to the degree that they introduce few additional complications to the Standard Model of physics and/or simultaneously solve other outstanding problems in fundamental physics, cosmology, and/or astrophysics.} that the missing piece is likely to be an as-yet undiscovered subatomic particle that pervades the universe (the ``dark matter particle''), although other solutions, such as alternative theories of gravity, are favored by some. It is clear that the discovery of the dark matter particle would be a Nobel-Prize-worthy achievement that would settle long-simmering debates and reveal much about the structure of the Universe from the smallest to the largest scales. 

Many searches for the particle begin from a guess at the manner in which the particle might interact with ordinary matter and light, followed by a search for the products of that interaction.  For instance, if the particle self-annihilates, then regions with large dark matter densities, such as the cores of stars or centers of galaxies, should ``glow'' with their annihilation products, such as neutrinos or $\gamma$ rays. Alternatively, if the dark matter particle has a non-zero scattering cross-section with ordinary particles, then their flux through a laboratory detector could be measured from the otherwise unexpected recoil of an atomic nucleus from a rare but inevitable collision with a dark matter particle \citep{DarkMatterDetection}.

On the other hand, the dark matter particle could be completely sterile (having no scattering cross section or self-annihilation processes involving ordinary matter or light), or there might be no dark matter particle at all.  In both cases, dark matter detection experiments would always come up empty, forever putting upper limits on certain parameters of proposed dark matter particles.

Similarly, SETI begins with a strong rationale: we know that intelligent life in the universe exists on Earth, and we seek to know more about its nature throughout the universe. Its form elsewhere, however, is uncertain.  Many arguments are made regarding its existence and detectability, but these are mostly made on philosophical grounds with weak empirical basis.  It is clear that the discovery of alien life would be a seminal achievement in science that would settle long-simmering debates and reveal much about the nature of life in the Universe. The discovery of intelligent alien life, then, would be an even more profound realization of the goal to understand the Universe and our place in it.

Many searches for intelligent alien life begin from a guess at the manner in which other civilizations might use matter and light, perhaps even in a deliberate attempt to be noticed, followed by a search for that effect.  For instance, if they transmit radio signals (as we do for communication or radar) then sufficiently powerful signals could be detected by our radio telescopes and would be unambiguously artificial in origin \citep{SETI,Tarter01}. Alternatively, if alien civilizations, being quite old, use very large amounts of energy harnessed from stars (for instance), this would be detectable as a dimunition of starlight\citep{Arnold05} or increase in infrared radiation \citep{dyson60}.

On the other hand, distant, intelligent alien life might not be detectable with any technology we will ever possess, or we may be unique.  In both cases, SETI would always come up empty, forever putting upper limits on certain parameters of proposed alien civilizations.

Although the analogy is not perfect (we have observational evidence for dark matter, but not for intelligent {\it extraterrestrial} species), it still shows that the disparity in funding between the two fields is not based on some fundamental difference in their plausibility, foundations, approach, or scientific rigor.  Other forces, such as the ``giggle factor,'' thus clearly play a role in SETI's exclusion from the national research portfolio.

\section*{Acknowledgements}

We thank the referee(s) for a constructive report that improved the structure and argumentation in this work. We thank John Johnson, Nia Imara, James Guillochon, Eric Kansa, and Adam Frank for their comments on drafts of this paper. 

J.T.W.\ first encountered many of the references and concepts in this paper via the work of Chanda Prescod-Weinstein and Zuleyka Zevallos. J.T.W.\ thanks Adrian Lucy, Alessondra Springmann, Arpita Roy, Alan Reyes, Scott Colby, James Howell, John Meier, Manuel Llinas, and Ryan Myers for useful ideas and citations. J.T.W.\ also thanks Garson O'Toole for hunting down the Clarke space elevator quote.

M.O.R.\ is grateful to SSHRC and the Vanier CGS program for supporting this research, and Simon Springer and the University of Victoria for affiliate support during fieldwork. M.O.R.\ thanks Robert McNees, Alice Gorman, Wayne Fife, Gary Catano, Lucianne Walkowicz, Frederick Scharmen, Taylor Genovese, Robin Whitaker, Mary Oman, Martin Pfeiffer, Tananarive Due, Matthew Daniels, and others for inspiring conversations and science fiction recommendations.

The Center for Exoplanets and Habitable Worlds is supported by the Pennsylvania State University, the Eberly College of Science, and the Pennsylvania Space Grant Consortium. This research was partially supported by Breakthrough Listen, part of the Breakthrough Initiatives sponsored by the Breakthrough Prize Foundation,\footnote{\url{http://www.breakthroughinitiatives.org}}; and by the Social Sciences and Humanities Research Council of Canada.

This research has made use of NASA's Astrophysics Data System.

\bibliography{references}

\begin{thebibliography}{}

\bibitem[Abrams, 2009]{ST:Reboot}
Abrams, J.~J. (2009).
\newblock {\em {Star} {Trek}}.
\newblock Paramount Pictures.

\bibitem[Achenbach, 2003]{achenbach2003captured}
Achenbach, J. (2003).
\newblock {\em Captured by Aliens: The Search for Life and Truth in a Very
  Large Universe}.
\newblock Kensington Publishing Corporation.

\bibitem[Anker, 2005]{anker_ecological_2005}
Anker, P. (2005).
\newblock The ecological colonization of space.
\newblock {\em Environmental History}, 10:239--268.

\bibitem[{Annis}, 1999]{Annis99a}
{Annis}, J. (1999).
\newblock {An astrophysical explanation for the ``great silence''.}
\newblock {\em Journal of the British Interplanetary Society}, 52:19--22.

\bibitem[{Arnold}, 2005]{Arnold05}
{Arnold}, L.~F.~A. (2005).
\newblock {Transit Light-Curve Signatures of Artificial Objects}.
\newblock {\em \apj}, 627:534--539.

\bibitem[Asimov, 1981]{asimov1981asimov}
Asimov, I. (1981).
\newblock {\em Asimov on Science Fiction}.
\newblock Doubleday.

\bibitem[Baker, 1988]{baker_alien_1988}
Baker, G. (1988).
\newblock {\em Alien Nation}.
\newblock 20th Century Fox.

\bibitem[Barr, 1992]{barr_feminist_1992}
Barr, M.~S. (1992).
\newblock {\em Feminist Fabulation: Space/Postmodern Fiction}.
\newblock University of Iowa Press, Iowa City.

\bibitem[Battaglia, 2006]{battaglia_e.t._2006}
Battaglia, D., editor (2006).
\newblock {\em E.{T}. {Culture}: {Anthropology} in {Outerspaces}}.
\newblock Duke University Press, Durham, N.C.

\bibitem[{Baum} et~al., 2011]{Baum11}
{Baum}, S.~D., {Haqq-Misra}, J.~D., and {Domagal-Goldman}, S.~D. (2011).
\newblock {Would contact with extraterrestrials benefit or harm humanity? A
  scenario analysis}.
\newblock {\em Acta Astronautica}, 68:2114--2129.

\bibitem[Berman, 1995]{ST:Voyager}
Berman, R. (1995).
\newblock {\em {Star} {Trek}: {Voyager}}.
\newblock CBS Television Distribution.

\bibitem[Billings, 2007]{billings_societal}
Billings, L. (2007).
\newblock Overview: Ideology, advocacy, and spaceflight---evolution of a
  cultural narrative.
\newblock In Dick, S.~J., editor, {\em Societal Impact of Spaceflight}, NASA
  SP, chapter~25, pages 483--499. National Aeronautics and Space
  Administration.

\bibitem[Blomkamp, 2009]{blomkamp_district_2009}
Blomkamp, N. (2009).
\newblock {\em District 9}.
\newblock TriStar Pictures.

\bibitem[Burton, 1993]{ST:TNG:SecondChances}
Burton, L. (1993).
\newblock {\em ``{\rm \uppercase{S}econd} {\rm \uppercase{C}hances}''}.
\newblock {\it Star Trek: The Next Generation}, Season 6, Episode 24, CBS
  Television Distribution.

\bibitem[Burton, 1996]{burton_mars_1996}
Burton, T. (1996).
\newblock {\em Mars Attacks!}
\newblock Warner Bros. Pictures.

\bibitem[{Cabrol}, 2016]{cabrol_alien_2016}
{Cabrol}, N.~A. (2016).
\newblock Alien mindscapes---a perspective on the search for extraterrestrial
  intelligence.
\newblock 16(9):661--676.

\bibitem[Capova, 2013]{capova_charming_2013}
Capova, K.~A. (2013).
\newblock {\em The Charming Science of the Other: The cultural analysis of the
  scientific search for life beyond earth.}
\newblock Phd thesis.

\bibitem[{Chyba} and {Hand}, 2005]{Chyba05}
{Chyba}, C.~F. and {Hand}, K.~P. (2005).
\newblock {Astrobiology: The Study of the Living Universe}.
\newblock {\em \araa}, 43:31--74.

\bibitem[{{\'C}irkovi{\'c}} and {Vukoti{\'c}}, 2008]{cirkovic08}
{{\'C}irkovi{\'c}}, M.~M. and {Vukoti{\'c}}, B. (2008).
\newblock {Astrobiological Phase Transition: Towards Resolution of Fermi's
  Paradox}.
\newblock {\em Origins of Life and Evolution of the Biosphere}, 38:535--547.

\bibitem[Clarke, 1951]{clarkeexploration}
Clarke, A. (1951).
\newblock {\em The exploration of space}.
\newblock Temple Press.

\bibitem[Clarke, 1953]{clarke_childhood}
Clarke, A. (1953).
\newblock {\em Childhood's End}.
\newblock Ballantine Books.

\bibitem[Clarke, 1962]{clarke_profiles}
Clarke, A. (1962).
\newblock {\em Profiles of the Future: An Inquiry Into the Limits of the
  Possible}.
\newblock Bantam science and mathematics. Harper \& Row.

\bibitem[{Clarke}, 1981]{ClarkeLaughing}
{Clarke}, A.~C. (1981).
\newblock The space elevator: ``thought experiment,'' or key to the universe?
\newblock {\em Advances in Earth Oriented Applied Space Technologies},
  1(1):39--48.

\bibitem[Cocconi and Morrison, 1959]{cocconi_searching_1959}
Cocconi, G. and Morrison, P. (1959).
\newblock Searching for interstellar communications.
\newblock {\em Nature}, 184(4690):844--846.

\bibitem[{Cocconi} and {Morrison}, 1959]{SETI}
{Cocconi}, G. and {Morrison}, P. (1959).
\newblock {Searching for Interstellar Communications}.
\newblock {\em \nat}, 184:844--846.

\bibitem[Collins, 2008]{collins_all_2008}
Collins, S.~G. (2008).
\newblock {\em All Tomorrow's Cultures: Anthropological Engagements with the
  Future}.
\newblock Berghahn Books.

\bibitem[Crenshaw, 1989]{crenshaw_demarginalizing_1989}
Crenshaw, K. (1989).
\newblock Demarginalizing the intersection of race and sex: A black feminist
  critique of antidiscrimination doctrine, feminist theory and antiracist
  politics.
\newblock {\em University of Chicago Legal Forum}, 1989:139.

\bibitem[Crenshaw, 1991]{crenshaw_mapping_1991}
Crenshaw, K. (1991).
\newblock Mapping the margins: Intersectionality, identity politics, and
  violence against women of color.
\newblock {\em Stanford Law Review}, 43(6):1241--1299.

\bibitem[Dean, 1998]{dean_aliens_1998}
Dean, J. (1998).
\newblock {\em Aliens in {America}: {Conspiracy} {Cultures} from {Outerspace}
  to {Cyberspace}}.
\newblock Cornell University Press, Ithaca, N.Y.

\bibitem[Denning, 2011]{denning_ten_2011}
Denning, K. (2011).
\newblock Ten thousand revolutions: conjectures about civilizations.
\newblock {\em Acta Astronautica}, 68(3):381--388.

\bibitem[Denning, 2013]{denning_impossible_2013}
Denning, K. (2013).
\newblock Impossible predictions of the unprecedented: Analogy, history, and
  the work of prognostication.
\newblock In Vakoch, D.~A., editor, {\em Astrobiology, History, and Society},
  Advances in Astrobiology and Biogeophysics, pages 301--312. Springer Berlin
  Heidelberg.

\bibitem[Dick, 2006]{dick_anthropology_2006}
Dick, S.~J. (2006).
\newblock Anthropology and the search for extraterrestrial intelligence: {An}
  historical view.
\newblock {\em Anthropology Today}, 22(2):3--7.

\bibitem[Dillon, 2012]{dillon_walking_2012}
Dillon, G.~L. (2012).
\newblock {\em Walking the clouds: an anthology of indigenous science fiction}.
\newblock University of Arizona Press, Tucson.

\bibitem[{Drake}, 1980]{DrakeEq}
{Drake}, F. (1980).
\newblock N is neither very small nor very large.
\newblock In {Papagiannis}, M.~D., editor, {\em {Strategies for the search for
  life in the universe : a joint session of Commissions 16, 40, and 44, held in
  Montreal, Canada, during the IAU General Assembly, 15 and 16 August, 1979}}.
  Dordrecht, Holland ; Boston : D.~Reidel Pub.~Co.~: sold and distributed in
  the U.S.A.~and Canada by Kluwer Boston, c1980.

\bibitem[{Drake}, 1961]{OZMA}
{Drake}, F.~D. (1961).
\newblock {Project OZMA}.
\newblock {\em Physics Today}, 14:40--46.

\bibitem[{Dyson}, 1960]{dyson60}
{Dyson}, F.~J. (1960).
\newblock {Search for Artificial Stellar Sources of Infrared Radiation}.
\newblock {\em Science}, 131:1667--1668.

\bibitem[Emmerich, 1996]{emmerich_independence_1996}
Emmerich, R. (1996).
\newblock {\em Independence Day}.
\newblock 20th Century Fox.

\bibitem[Finney and Bentley, 2014]{vakoch_tale_2014}
Finney, B. and Bentley, J. (2014).
\newblock A {Tale} of {Two} {Analogues}: {Learning} at a {Distance} from the
  {Ancient} {Greeks} and {Maya} and the {Problem} of {Deciphering}
  {Extraterrestrial} {Radio} {Transmissions}.
\newblock In Vakoch, D.~A., editor, {\em Archaeology, {Anthropology}, and
  {Interstellar} {Communication}}, pages 65--78. National Aeronautics and Space
  Administration, Office of Communications, History Program Office, Washington,
  DC.

\bibitem[{Garber}, 1999]{Garber99}
{Garber}, S.~J. (1999).
\newblock {Searching for Good Science --- The Cancellation of NASA's SETI
  Program}.
\newblock {\em Journal of the British Interplanetary Society}, 52:3--12.

\bibitem[Genovese, 2017]{genovese_new_2017}
Genovese, T.~R. (2017).
\newblock The new right stuff: Social imaginaries of outer space and the
  capitalist accumulation of the cosmos.
\newblock Ma thesis.

\bibitem[{Gertz}, 2016]{Gertz_METI}
{Gertz}, J. (2016).
\newblock {Reviewing METI: A Critical Analysis of the Arguments}.
\newblock {\em arXiv:1605.05663}.

\bibitem[{Gertz}, 2017]{Gertz17}
{Gertz}, J. (2017).
\newblock {Post-Detection SETI Protocols {\&} METI: The Time Has Come To
  Regulate Them Both}.
\newblock {\em arXiv:1701.08422}.

\bibitem[Gilman, 2015]{gilman_dark_2015}
Gilman, C.~I. (2015).
\newblock {\em Dark Orbit}.
\newblock Tor, New York.

\bibitem[Gorman, 2009]{gorman_cultural_2009}
Gorman, A. (2009).
\newblock Cultural landscape of space.
\newblock In Darrin, A.~G. and O'Leary, B.~L., editors, {\em Handbook of
  {Space} {Engineering}, {Archaeology}, and {Heritage}}, pages 335--346. CRC
  Press.

\bibitem[{Gott}, 1993]{Gott93}
{Gott}, III, J.~R. (1993).
\newblock {Implications of the Copernican principle for our future prospects}.
\newblock {\em \nat}, 363:315--319.

\bibitem[Graeber, 2015]{graeber_utopia_2015}
Graeber, D. (2015).
\newblock {\em The utopia of rules: on technology, stupidity, and the secret
  joys of bureaucracy}.
\newblock Melville House.
\newblock {OCLC}: 908261537.

\bibitem[{Gray}, 2015]{FermiParadox}
{Gray}, R.~H. (2015).
\newblock {The Fermi Paradox Is Neither Fermi's Nor a Paradox}.
\newblock {\em Astrobiology}, 15:195--199.

\bibitem[Grewell, 2001]{grewell2001colonizing}
Grewell, G. (2001).
\newblock Colonizing the universe: Science fictions then, now, and in the
  (imagined) future.
\newblock {\em Rocky Mountain Review of Language and Literature}, pages 25--47.

\bibitem[{Haqq-Misra}, 2016]{Haqq16}
{Haqq-Misra}, J. (2016).
\newblock {The Transformative Value of Liberating Mars}.
\newblock {\em New Space}, 4:64--67.

\bibitem[Haraway, 2013]{haraway_sf:_2013}
Haraway, D.~J. (2013).
\newblock {SF}: {Science} {Fiction}, {Speculative} {Fabulation}, {String}
  {Figures}, {So} {Far}.
\newblock {\em Ada: A Journal of Gender, New Media, and Technology,
  doi:10.7264/N3KH0K81}, (3).

\bibitem[Harrison et~al., 2000]{harrison_role_2000}
Harrison, A., Billingham, J., Dick, S.~J., Finney, B., Michaud, M.~A., Tarter,
  D.~E., and Vakoch, D.~A. (2000).
\newblock The role of the social sciences in {SETI}.
\newblock In Tough, A., editor, {\em When {SETI} succeeds: the impact of
  high-information contact}, volume~1, pages 71--85. Foundation For the Future,
  Bellevue, Wash.

\bibitem[Harrison, 2005]{harrison_overcoming_2005}
Harrison, A.~A. (2005).
\newblock Overcoming the image of little green men: Astrosociology and {SETI}.
\newblock California Sociological Association Conference.

\bibitem[{Hart}, 1975]{hart75}
{Hart}, M.~H. (1975).
\newblock {Explanation for the Absence of Extraterrestrials on Earth}.
\newblock {\em \qjras}, 16:128.

\bibitem[Heinlein, 1965]{heinlein_freehold}
Heinlein, R. (1965).
\newblock {\em Farnham's Freehold}.
\newblock Signet books. New American Library.

\bibitem[Herzing, 2014]{herzing_profiling_2014}
Herzing, D.~L. (2014).
\newblock Profiling nonhuman intelligence: An exercise in developing unbiased
  tools for describing other ''types'' of intelligence on earth.
\newblock {\em Acta astronautica.}, 94(2):676--680.

\bibitem[Imarisha and Brown, 2015]{imarisha_octavias_2015}
Imarisha, W. and Brown, A.~M., editors (2015).
\newblock {\em Octavia's {Brood}: {Science} {Fiction} {Stories} from {Social}
  {Justice} {Movements}.}
\newblock A K Press, Oakland.

\bibitem[Irvine and Gal, 2000]{kroskrity_language_2000}
Irvine, J.~T. and Gal, S. (2000).
\newblock Language ideology and linguistic differentiation.
\newblock In Kroskrity, P., editor, {\em Regimes of language: ideologies,
  polities, and identities}. School of American Research Press, Santa Fe.

\bibitem[Lambourne et~al., 1990]{lambourne_close_1990}
Lambourne, R., Shallis, M., and Shortland, M. (1990).
\newblock {\em Close encounters?: science and science fiction}.
\newblock Hilger.

\bibitem[Le~Guin, 1974]{le_guin_dispossessed_1974}
Le~Guin, U.~K. (1974).
\newblock {\em The Dispossessed}.
\newblock HarperCollins, New York.

\bibitem[Le~Guin, 2004]{le_guin_wave_2004}
Le~Guin, U.~K. (2004).
\newblock {\em The wave in the mind: talks and essays on the writer, the
  reader, and the imagination}.
\newblock Shambhala ; Distributed in the United States by Random House.
\newblock {OCLC}: 52418183.

\bibitem[Lem, 1970]{stanislaw_lem_solaris_1970}
Lem, S. (1970).
\newblock {\em Solaris}.
\newblock Harcourt, San Diego.

\bibitem[Lempert, 2014]{lempert_decolonizing_2014}
Lempert, W. (2014).
\newblock Decolonizing {Encounters} of the {Third} {Kind}: {Alternative}
  {Futuring} in {Native} {Science} {Fiction} {Film}.
\newblock {\em VAR Visual Anthropology Review}, 30(2):164--176.

\bibitem[{Lingam} and {Loeb}, 2017]{Loeb17}
{Lingam}, M. and {Loeb}, A. (2017).
\newblock {Fast Radio Bursts from Extragalactic Light Sails}.
\newblock {\em \apjl}, 837:L23.

\bibitem[Lockhart, 2015]{lockhart_nerd/geek_2015}
Lockhart, E.~A. (2015).
\newblock {\em Nerd/{Geek} {Masculinity}: {Technocracy}, {Rationality}, and
  {Gender} in {Nerd} {Culture}'s {Countermasculine} {Hegemony}}.
\newblock Thesis.

\bibitem[Lowe, 1996]{ehrich_van_lowe_homeboys_1996}
Lowe, E.~V. (1996).
\newblock {\em Homeboys in Outer Space}.
\newblock UPN.

\bibitem[{Marrod{\'a}n Undagoitia} and {Rauch}, 2016]{DarkMatterDetection}
{Marrod{\'a}n Undagoitia}, T. and {Rauch}, L. (2016).
\newblock {Dark matter direct-detection experiments}.
\newblock {\em Journal of Physics G Nuclear Physics}, 43(1):013001.

\bibitem[{McCray}, 2012]{mccray_visioneers:_2012}
{McCray}, P. (2012).
\newblock {\em The Visioneers: How a Group of Elite Scientists Pursued Space
  Colonies, Nanotechnologies, and a Limitless Future}.
\newblock Princeton University Press.

\bibitem[Messeri, 2017]{messeri_resonant_2017}
Messeri, L. (2017).
\newblock Resonant worlds: Cultivating proximal encounters in planetary
  science.
\newblock {\em American Ethnologist}, 44(1):131--142.

\bibitem[Milburn, 2010]{Milburn:2010:MFS}
Milburn, C. (2010).
\newblock Modifiable futures: Science fiction at the bench.
\newblock {\em j-ISIS}, 101(3):560--569.

\bibitem[Morgan, 2014]{morgan_rise_2014}
Morgan, A. (2014).
\newblock The {Rise} of the {Geek}: {Exploring} {Masculine} {Identity} in {The}
  {Big} {Bang} {Theory}.

\bibitem[Nama, 2008]{nama_black_2008}
Nama, A. (2008).
\newblock {\em Black Space: Imagining Race in Science Fiction Film}.
\newblock University of Texas Press.

\bibitem[{National Aeronautics and Space Administration},
  2015]{NASA_Astrobiology_2015}
{National Aeronautics and Space Administration} (2015).
\newblock {\em {NASA} Astrobiology Strategy 2015}.
\newblock NASA.

\bibitem[Olson, 2010]{olson_american_2010}
Olson, V.~A. (2010).
\newblock {\em American Extreme: An Ethnography of Astronautical Visions and
  Ecologies}.
\newblock Phd thesis.

\bibitem[Oman-Reagan, 2015a]{oman-reagan_social_2015}
Oman-Reagan, M.~P. (2015a).
\newblock The social lives of plants, in space.
\newblock {\em Astrosociological Insights}, 4(2):4--8.

\bibitem[Oman-Reagan, 2015b]{oman-reagan_unfolding_2015}
Oman-Reagan, M.~P. (2015b).
\newblock Unfolding the {Space} {Between} {Stars}: {Anthropology} of the
  {Interstellar}.
\newblock In {\em 114th Annual Meeting of the American Anthropological
  Association}. American Anthropological Association.

\bibitem[{O'Neill}, 1974]{ONeill}
{O'Neill}, G.~K. (1974).
\newblock {The Colonization of Space}.
\newblock {\em Physics Today}, 27(9):32--40.

\bibitem[Redfield, 2002]{redfield_half-life_2002}
Redfield, P. (2002).
\newblock The half-life of empire in outer space.
\newblock {\em Social Studies of Science}, 32(5):791--825.

\bibitem[Robinson, 1993]{RedMars}
Robinson, K. (1993).
\newblock {\em Red Mars}.
\newblock Bantam Spectra science fiction. Bantam Books.

\bibitem[Roddenberry, 1966]{ST:TOS}
Roddenberry, G. (1966).
\newblock {\em {Star} {Trek}}.
\newblock CBS Television Distribution.

\bibitem[{Saberhagen}, 1967]{Berserker}
{Saberhagen}, F. (1967).
\newblock {\em Berserker}.
\newblock Ballantine, New York, NY, USA.

\bibitem[Sagan, 1985a]{contact}
Sagan, C. (1985a).
\newblock {\em Contact: a novel}.
\newblock Simon and Schuster.

\bibitem[Sagan, 1985b]{sagan1985cosmos}
Sagan, C. (1985b).
\newblock {\em Cosmos}.
\newblock Cognitive systems monographs. Ballantine Books.

\bibitem[Sagan and Druyan, 1994]{PaleBlueDot}
Sagan, C. and Druyan, A. (1994).
\newblock {\em Pale Blue Dot: A Vision of the Human Future in Space}.
\newblock Random House Publishing Group.

\bibitem[{Shostak}, 2011]{DrakeL}
{Shostak}, S. (2011).
\newblock {L: How Long Do They Last?}
\newblock In {Shuch}, H.~P., editor, {\em Searching for Extraterrestrial
  Intelligence}, page 451.

\bibitem[Sontag, 1966]{SontagCamp}
Sontag, S. (1966).
\newblock {\em Against Interpretation: And Other Essays}.
\newblock Delta book. Picador.

\bibitem[Spielberg, 1977]{spielberg_close_1977}
Spielberg, S. (1977).
\newblock {\em Close Encounters of the Third Kind}.
\newblock Columbia Pictures.

\bibitem[Swain and Nieli, 2003]{Hart_White_Separatism}
Swain, C. and Nieli, R. (2003).
\newblock {\em Contemporary Voices of White Nationalism in America}.
\newblock Cambridge University Press.

\bibitem[{Tarter}, 2001]{Tarter01}
{Tarter}, J. (2001).
\newblock {The Search for Extraterrestrial Intelligence (SETI)}.
\newblock {\em \araa}, 39:511--548.

\bibitem[Thomas, 2000]{thomas_dark_2000}
Thomas, S.~R. (2000).
\newblock {\em Dark matter: a century of speculative fiction from the African
  diaspora}.
\newblock Warner Books.

\bibitem[{Tipler}, 1980]{tipler80}
{Tipler}, F.~J. (1980).
\newblock {Extraterrestrial intelligent beings do not exist}.
\newblock {\em \qjras}, 21:267--281.

\bibitem[Tsing, 2014]{tsing_strathern_2014}
Tsing, A.~L. (2014).
\newblock Strathern beyond the human: Testimony of a spore.
\newblock {\em Theory, Culture \& Society}, 31(2):221--241.

\bibitem[Vakoch, 1998]{vakoch_dialogic_1998}
Vakoch, D.~A. (1998).
\newblock The dialogic model: representing human diversity in messages to
  extraterrestrials.
\newblock {\em Acta Astronautica}, 42:705--710.

\bibitem[Valentine, 2012]{valentine_exit_2012}
Valentine, D. (2012).
\newblock Exit strategy: Profit, cosmology, and the future of humans in space.
\newblock {\em Anthropological Quarterly}, 85(4):1045--1067.

\bibitem[Watts, 2006]{watts_blindsight_2006}
Watts, P. (2006).
\newblock {\em Blindsight}.
\newblock Tor, New York.

\bibitem[Wells, 1898]{wells1898war}
Wells, H. (1898).
\newblock {\em The War of the Worlds}.
\newblock B. Tauchnitz.

\bibitem[{Wright}, 2017]{PITS}
{Wright}, J.~T. (2017).
\newblock {Prior Indigenous Technological Species}.
\newblock {\em arXiv:1704.0763; International Journal of Astrobiology {\it in
  production}}.

\bibitem[{Wright} et~al., 2014]{GHAT1}
{Wright}, J.~T., {Mullan}, B., {Sigurdsson}, S., and {Povich}, M.~S. (2014).
\newblock {The {\^G} Infrared Search for Extraterrestrial Civilizations with
  Large Energy Supplies. I. Background and Justification}.
\newblock {\em \apj}, 792:26.

\bibitem[{Yurtsever} and {Wilkinson}, 2015]{Yurtsever15}
{Yurtsever}, U. and {Wilkinson}, S. (2015).
\newblock {Limits and Signatures of Relativistic Spaceflight}.
\newblock {\em arXiv:1503.05845}.

\bibitem[Zuckerman and Hart, 1995]{zuckerman1995extraterrestrials}
Zuckerman, B. and Hart, M. (1995).
\newblock {\em Extraterrestrials: Where Are They?}
\newblock Cambridge University Press.

\end{thebibliography}
\end{document}